%% file: main.tex
\documentclass[runningheads]{llncs}

\usepackage{graphicx}
\usepackage{subcaption}
\usepackage[hidelinks]{hyperref}
\usepackage{amsmath}
\usepackage{booktabs} 
\usepackage{multirow}  
\usepackage{makecell}  
\usepackage{bigstrut}

\begin{document}

\title{CardioSpectrum: Comprehensive Myocardium Motion Analysis with 3D Deep Learning and Geometric Insights}

\author{Shahar Zuler\inst{1} \and
Shai Tejman-Yarden\inst{2,3,4} \and
Dan Raviv\inst{1}}

\authorrunning{S. Zuler et al.}

\institute{Tel Aviv University, Tel Aviv, Israel \\
\email{shahar.zuler@gmail.com} \email{darav@tauex.tau.ac.il} 
\and 
Faculty of Medicine, Tel Aviv University, Tel Aviv, Israel.\\
\inst{3}The Edmond J. Safra International Congenital Heart Center, Sheba Medical Center, Ramat Gan, Israel.\\
\inst{4}The Engineering Medical Research Laboratory, Sheba Medical Center, Ramat Gan, Israel. \\
\email{tegmanya@gmail.com}}

\maketitle

\markboth{Zuler et al.}{CardioSpectrum: Comprehensive Myocardium Motion Analysis}

\begin{abstract}
\input{00_abstract}
\end{abstract}

\begin{figure}[h]
    \begin{subfigure}{0.5\textwidth} 
        \includegraphics[width=\linewidth]{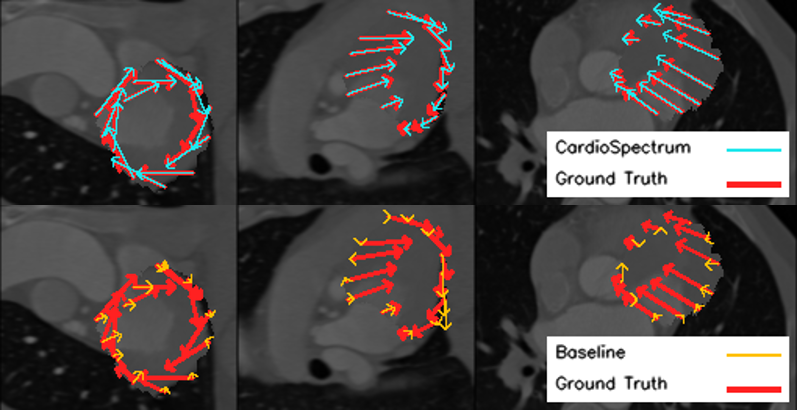}
        \caption{CardioSpectrum effectively captures cardiac tangential movement (top), outperforming traditional methods that predominantly focus on local features, which often lead to inadequate results due to the aperture problem (bottom).}
        \label{fig:teaser_sample}
    \end{subfigure}
    \hfill
    \begin{subfigure}{0.45\textwidth} 
        \includegraphics[width=\linewidth]{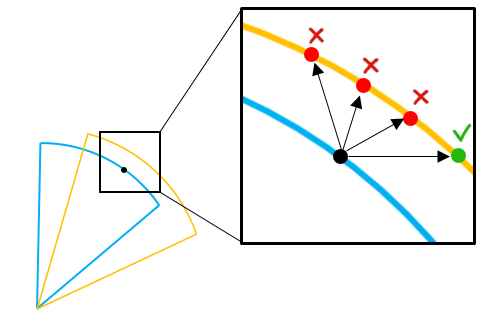}
        \caption{The aperture problem: Global comprehension of geometry is essential to effectively capture tangential movements.}
        \label{fig:aperature_prob}
    \end{subfigure}    
    \caption{}
    \label{fig:teaser}
\end{figure}

\section{Introduction}
\input{01_introduction}

\section{Motivation}
\input{02_motivation}

\section{The Aperture Problem}
\input{03_the_aperture_problem}

\section{Method}
\input{04_methods}

\section{Experiments and Results}
\input{05_experiments_and_results}

\section{Discussion and Conclusion}
\input{06_discussion_and_conclusions}

\subsubsection*{Acknowledgments.} This work is partially funded by the Zimin Institute for Engineering Solutions Advancing Better Lives.

\subsubsection*{Disclosure of Interests.}
The authors have no competing interests to declare that are
relevant to the content of this article.

\bibliographystyle{splncs04} 
\bibliography{08_references}

\clearpage

\section*{\centering Supplementary Material}
\addcontentsline{toc}{section}{Supplementary Material}
\markboth{Supplementary Material}{Supplementary Material}
\input{09_supplemetary_material}


\end{document}

%% file: 00_abstract.tex
The ability to map left ventricle (LV) myocardial motion using computed tomography angiography (CTA) is essential to diagnosing cardiovascular conditions and guiding interventional procedures. Due to their inherent locality, conventional neural networks typically have difficulty predicting subtle tangential movements, which considerably lessens the level of precision at which myocardium three-dimensional (3D) mapping can be performed. Using 3D optical flow techniques and Functional Maps (FMs), we present a comprehensive approach to address this problem. FMs are known for their capacity to capture global geometric features, thus providing a fuller understanding of 3D geometry. As an alternative to traditional segmentation-based priors, we employ surface-based two-dimensional (2D) constraints derived from spectral correspondence methods. Our 3D deep learning architecture, based on the ARFlow model, is optimized to handle complex 3D motion analysis tasks. By incorporating FMs, we can capture the subtle tangential movements of the myocardium surface precisely, hence significantly improving the accuracy of 3D mapping of the myocardium. The experimental results confirm the effectiveness of this method in enhancing myocardium motion analysis. This approach can contribute to improving cardiovascular diagnosis and treatment.
Our code and additional resources are available at: \url{https://shaharzuler.github.io/CardioSpectrumPage} 

%% file: 01_introduction.tex
The accurate mapping of LV myocardium movement is a critical feature of cardiac CT analysis. Understanding the 3D motion of the chambers and vessels of the heart is essential to diagnosing cardiovascular conditions and guiding medical interventions. However, this task poses several challenges, including the need to capture the precise tangential movements of the myocardium.

Traditional neural networks (NNs) have shown promise in analyzing cardiac movement from imaging data \cite{duchateau2020machine} through the use of designated architectures \cite{Meng2022MulViMotionS3}\cite{morales2019implementation}\cite{qin2023fsdiffreg}\cite{xu2023importance}, optimization schemes \cite{dou2023gsmorph}, priors and constraints \cite{yan2019cine}\cite{hanania2023pcmc}\cite{balakrishnan2019voxelmorph}\cite{Morales2021}. However, in addition to the scarcity of annotated high-resolution imagery, they also often suffer from the locality problem, since they struggle to predict tangential movements accurately (as depicted in figure \ref{fig:teaser_sample}). While they excel at minimizing photometric energy to capture radial movements of the myocardium, their inherent locality limits their ability to comprehensively model subtle tangential movements. This limitation is referred to as \textbf{the aperture problem}, as illustrated in figure \ref{fig:aperature_prob}.

To address the challenge of capturing both radial and tangential components of myocardium motion, we propose a holistic approach that integrates DL techniques with spectral methods, and specifically FMs~\cite{ovsjanikov2012functional}. FMs, initially developed for non-rigid shape matching, operate by projecting functions from different shapes onto the eigenfunctions of the Laplacian Beltrami Operator (LBO)~\cite{lbos}. This LBO projection effectively captures the intrinsic geometry of the shapes, reducing the dimensionality of the problem and improving solution accuracy.
By considering the LV surface as a manifold, we leverage FMs' global perception for understanding myocardium motion.

Our comprehensive approach as detailed in this paper includes a 3D DL architecture and the incorporation of spectral methods. We leverage 2D constraints derived from spectral correspondence methods to significantly enhance the accuracy of surface mapping. These constraints provide a more robust foundation compared to traditional segmentation priors by precisely specifying the pixel correspondences between timesteps.

The following sections detail our approach, and provide insights into each component that contributes to the accuracy and inclusiveness of our myocardium 3D mapping technique. We believe that this approach represents a significant step forward in cardiac CT analysis, by addressing the challenges associated with capturing both radial and tangential myocardium movements accurately.

%% file: 02_motivation.tex
Understanding cardiovascular health extends beyond the study of electrical signals and anatomical structures. It hinges on comprehending the intricate dynamics of the myocardium—the muscular heart wall responsible for contractions—and its movement throughout the cardiac cycle. Accurate measurement of myocardium behavior is not just an academic aspiration, but rather is pivotal to patient well-being and the advancement of cardiovascular research. 

The heart operates in a rhythmic cycle, where each beat relies on the orchestrated movements of the myocardium. This muscular layer governs the heart's efficiency in pumping blood. Perturbations in myocardial movement can manifest as diverse cardiovascular diseases, ranging from heart attacks to cardiomyopathies.

Despite advances in technology, inaccuracies in myocardium movement extraction persist. 
These inaccuracies can result in incorrect diagnoses, ineffective treatment plans, and slow progress in understanding cardiac pathophysiology.

In the field of medical imaging, experts often work with 2D slices of the heart, which requires mental integration to understand its 3D structure and movements. This process can be challenging. By contrast, computer algorithms excel in handling 3D or 3D+time (3D+t) data, and can provide a holistic view of the heart's structure and motion. When analyzing a 3D+t cardiac CT scan, for example, these algorithms generate a high-resolution 3D movement field that can be translated into deformation, velocity, and strain. These data can be represented visually, thus enhancing interpretability for medical professionals.

Accurate myocardium movement extraction and the application of 3D imaging and computer vision do not only constitute medical advances but are also holistic endeavors that can revolutionize cardiovascular health. These pursuits are not only likely to generate scientific knowhow but also tangible, life-improving benefits for individuals worldwide.

%% file: 03_the_aperture_problem.tex
In addressing 3D optical flow within a localized window, our goal is to minimize:
\begin{equation} \label{eq:1}
    || I_x u + I_y v + I_z w + I_t ||
\end{equation}
where $u$, $v$, and $w$, the optimized parameters, correspond to the optical flow vector elements indicating motion along the $x$, $y$, and $z$ axes, whereas $I_x$, $I_y$, $I_z$ and $I_t$ denote the spatial and temporal derivatives of pixel intensities.

Consider a 3D image characterized by pixel intensities, where we assume that these intensities define a level set within a small 3D window. If the level set undergoes motion, the motion vector can be decomposed into components along different axes. These include a component along the axis $\mathcal{N}$, normal to the level set, and components along axes $t_1$ and $t_2$, lying in a plane perpendicular to $\mathcal{N}$ and mutually perpendicular. 
The objective term to minimize in this scenario can be expressed as:
\begin{equation} 
    || I_\mathcal{N} f_\mathcal{N} + I_{t_1} f_{t_1} + I_{t_2} f_{t_2} + I_t ||
\end{equation}
As $I_{t_1}$ and $I_{t_2}$ vanish due to the motion within the level set plane, the optimization problem simplifies to:
\begin{equation}
    || I_\mathcal{N} f_\mathcal{N} + I_t ||
\end{equation}
This reduction implies that the solution to the 3D optical flow problem only considers the radial component, $f_\mathcal{N}$, and disregards the tangential components, $f_{t_1}$ and $f_{t_2}$. In essence, it exclusively addresses motion along the axis normal to the level set, leaving the tangential components unconstrained. This is called the aperture problem.

Global approaches like Horn and Schunk \cite{HORN1981185} and Convolutional Neural Networks (CNNs) have been suggested but remain curtailed by the aperture problem. Despite CNNs' broader perspective, their local context makes them prone to this issue. In cardiac CT imaging, this limitation is a major impediment because CNNs may fail to accurately estimate tangential motion on the dynamic LV myocardium surface due to their localized receptive fields.

\subsubsection{Hypothesis:}
Introducing constraints on the level set may enhance optical flow evaluation around the level set, thus offering a potential approach to addressing the aperture problem.

%% file: 04_methods.tex
To address the limitations of a naive 3D optical flow approach in predicting the tangential component, we first present the global geometric solution. We then integrate the results of a global geometric solution as a constraint into the 3D voxel-based solution. This fusion yields a network that can simultaneously produce a dense flow field while correctly predicting the tangential components. Here, we define the LV surface as encompassing the outer boundary of the myocardium. In practice, this boundary constitutes the surface surrounding the union of the LV myocardium and the LV blood cavity.

\begin{figure}[h] 
    \centering
        \includegraphics[width=0.95\linewidth]{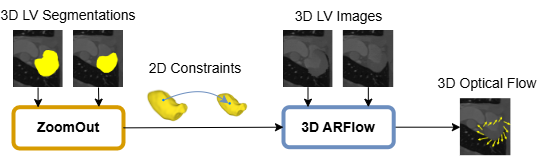}
        \caption{CardioSpectrum Architecture: The NN analyzes 3D image pairs of cardiac cycle timesteps, incorporating 2D constraints from ZoomOut. These constraints, derived from segmentations converted into meshes, result in a 3D optical flow.}
        \label{fig:sub7}
    \label{fig:archi}
\end{figure}
\subsection{Global Constraints from Functional Maps}
\input{04_methods_ch/04_b_2D_constraints}
\subsection{3D Optical Flow Network}
\input{04_methods_ch/04_a_3D_flow}
\subsection{Accuracy and Correspondence Measures}
\label{sec:accuracy}
\input{04_methods_ch/04_c_measures}
\subsection{Implementation details}
\input{04_methods_ch/04_d_implementation_details}

%% file: 04_methods_ch/04_b_2D_constraints.tex
To extract LV surface constraints, we use the ZoomOut method~\cite{melzi2019zoomout}, which iteratively evaluates the FM matrix, transitioning between spatial and spectral domains. This process employs a progressive zooming-out strategy, gradually expanding the dimensionality of the functional mapping and refining point correspondences at each step, establishing correspondences between triangle mesh representations of two timesteps, derived from their smoothed LV segmentations.

We map these surface constraints to a 3D grid by calculating an array of 3D displacements, $\textbf{D}$, by subtracting corresponding vertex locations on the meshes. Using rounded vertex locations as indices, we fill a 3D grid-based flow field with sparse flow values from $\textbf{D}$. Constraints are then interpolated into each voxel of the original unsmoothed segmentation map hull, providing the final constraints for the 3D optical flow network.

%% file: 04_methods_ch/04_a_3D_flow.tex
Our 3D optical flow network is built upon the ARFlow~\cite{Liu2020LearningBA} training methodology and recommended model architecture, which employs a PWC-Net-based model ~\cite{sun2018pwc}. Adapted for 3D imaging as outlined in~\cite{lifshitz2021unsupervised}, our model integrates essential components including a feature pyramid extractor, cost volume, a flow estimator, and a refinement context network. 
The core objective of our 3D optical flow network is expressed in the following equation:
\begin{equation}
\mathcal{L} = \lambda_{\mathcal{L}_{1}} \cdot \mathcal{L}_{1} + \lambda_{ssim} \cdot \mathcal{L}_{ssim} + \lambda_{constraints} \cdot \mathcal{L}_{constraints}
\end{equation}
\paragraph{$\mathcal{L}_{1}$ and $\mathcal{L}_{ssim}$ Losses:}
The $\mathcal{L}_{1}$ loss captures the pixel-wise intensity disparities, ensuring accurate matching between the target and warped source images. $\mathcal{L}_{ssim}$ \cite{1284395} focuses on preserving structural details and perceptual quality during optimization.
\paragraph{Constraints Loss:}
Our network integrates the constraints obtained from the FM pipeline operating on the LV surface. This constraints loss acts as a penalty term that incentivizes the predicted flows at constraint locations to match closely with the provided constraints:
\begin{equation}
\mathcal{L}_{constraints} = \sum_{i,j,k \in \text{segmentation hull}} \lVert \hat{\text{flow}}(i,j,k) - \text{constraint}(i,j,k) \rVert_1
\end{equation}
Here, $\hat{\text{flow}}(i,j,k)$ and $\text{constraint}(i,j,k)$ are the predicted flow and the provided constraint values at voxel locations $(i,j,k)$ within the segmentation hull. \\

Thus overall, our network is trained with a balanced combination of $\mathcal{L}_{1}$, $\mathcal{L}_{ssim}$, and $\mathcal{L}_{constraints}$, each weighted by the corresponding hyperparameters $\lambda_{\mathcal{L}_{1}}$, $\lambda_{ssim}$, and $\lambda_{constraints}$. 
These $\lambda$ values form vectors, where each element of the vectors corresponds to a layer in the feature pyramid of our network. 

\subsubsection{Optimization}
Our optimization scheme involves two stages. First, we pre-train our model on a large motion-based dataset from raw cardiac 4DCT scans. Next, we fine-tune the model for each individual pair using $\mathcal{L}{1}$ and $\mathcal{L}{ssim}$, followed by further optimization with $\mathcal{L}_{constraints}$.

%% file: 04_methods_ch/04_c_measures.tex
To assess the accuracy of the predicted flow values, we compared them to the ground truth flow values using the following metrics:
\subsubsection{Mean End-Point Error (mEPE)} quantifies the distance between the endpoints obtained from the predicted flow and the ground truth flow. It is defined as follows:
\begin{equation}
    mEPE(\hat{\text{flow}}, \text{flow}) = \frac{1}{N}\sum_{i,j,k} || \hat{\text{flow}}({i,j,k}) - \text{flow}({i,j,k}) || _2
\end{equation}
where $N$ denotes the number of voxels, $\hat{\text{flow}}$ represents the predicted flow, $\text{flow}$ denotes the ground truth flow, and $i$, $j$, $k$ are indices corresponding to voxels within a defined region.

We calculate the mEPE using flow values exclusively from voxels contained within the segmentation of the myocardium.
In addition, we compute the mEPE for the radial, longitudinal, and circumferential flow components separately, as illustrated in the supplementary material. We determine the longitudinal axis using the principal component of the LV segmentation map.

For the voxels contained in the LV segmentation hull, we separately compute the mEPE for the $locally$-$tangential$ and $locally$-$radial$ flow components, as illustrated in the supplementary material.

\subsubsection{Mean Angular Error}
Quantifies the dissimilarity between the predicted and the ground truth flow directions at each point in our analysis by computing the mean angle between the predicted and the true flow vectors.
\begin{equation}
    \overline{E}_{angular}(\hat{\text{flow}}, \text{flow}) = \frac{1}{N}\sum_{i,j,k} \left| \cos^{-1}\left(\frac{\hat{\text{flow}}(i,j,k) \cdot \text{flow}(i,j,k)}{\| \hat{\text{flow}}(i,j,k) \| _2 \| \text{flow}(i,j,k) \| _2}\right) \right|
\end{equation}
Here, $\cdot$ denotes the dot product between the two vectors. The other notations are identical to those described above.

%% file: 04_methods_ch/04_d_implementation_details.tex
We optimized our PyTorch \cite{NEURIPS2019_9015} network with seven feature pyramid layers, using the Adam optimizer~\cite{kingma2014adam} with a weight decay of $10^{-6}$, $\alpha=10^{-4}$, $\beta_1=0.9$, $\beta_2=0.999$, and $\epsilon=10^{-7}$. Training involved $2500$ unconstrained iterations and an additional $2500$ iterations with constraints. Losses for pyramid layers were weighted $\lambda_{ph}=3$, $\lambda_{ssim}=1$ and $\lambda_{constraints}=100$.
For ZoomOut preprocessing, the segmentation underwent morphological closing (kernel size 5) and opening (kernel size 7), followed by conversion to a triangle mesh using marching cubes \cite{lorensen1998marching} from \cite{scikit-learn}. Mesh smoothing used spectral filtering via the first 120 eigenfunctions of the mesh's LBOs \cite{lbos}.
ZoomOut initiated with $3$ eigenvectors and $3$ eigenvalues on each mesh, incorporating 50 Wave Kernel Signature (WKS) descriptors \cite{Aubry2011TheWK} over $130$ iterations. The entire optimization on a single Quadro RTX 8000 GPU took approx. 240 minutes and required 1900 MB of GPU memory.

%% file: 05_experiments_and_results.tex
Our experimental evaluations consisted of in-depth assessments of the efficacy of our method through optimization on a synthetic dataset enriched with ground truth annotations for LV 3D optical flow. The dataset consisted of 300 unique scan pairs of systole and diastole, each exhibiting torsion ranging equally from 0 to approximately 35 degrees and distributed across various torsion centers. The data distribution and deformation process are described in \cite{zuler2024synthetic}. The scans were taken from various sources \cite{gao2023bayeseg}\cite{zhuang2018multivariate}\cite{luo2022mathcal}\cite{wu2022minimizing}\cite{osirix}.

We compared the performance of our method against two baseline approaches. The first baseline, here termed "3D ARFlow", corresponds to the network before optimization with additional constraint flow, akin to \cite{lifshitz2021unsupervised}. The second baseline, akin to \cite{Morales2021}\cite{balakrishnan2019voxelmorph}, involved 3D ARFlow with an additional anatomical loss that encouraged maximal overlap between the predicted and ground truth segmentations. Both baselines were trained for 2500 iterations.

We observed that optimal outcomes were not always achieved at the point of convergence in iterations. This is a recognized issue, which is typically addressed through the optimization of hyperparameters. In our study, we halted the process based on performance against a deliberately introduced deformation. Below, we highlight the main results.
\subsection{Results}
\begin{figure}
    \centering
    \begin{subfigure}{0.34\textwidth} 
        \includegraphics[width=\linewidth]{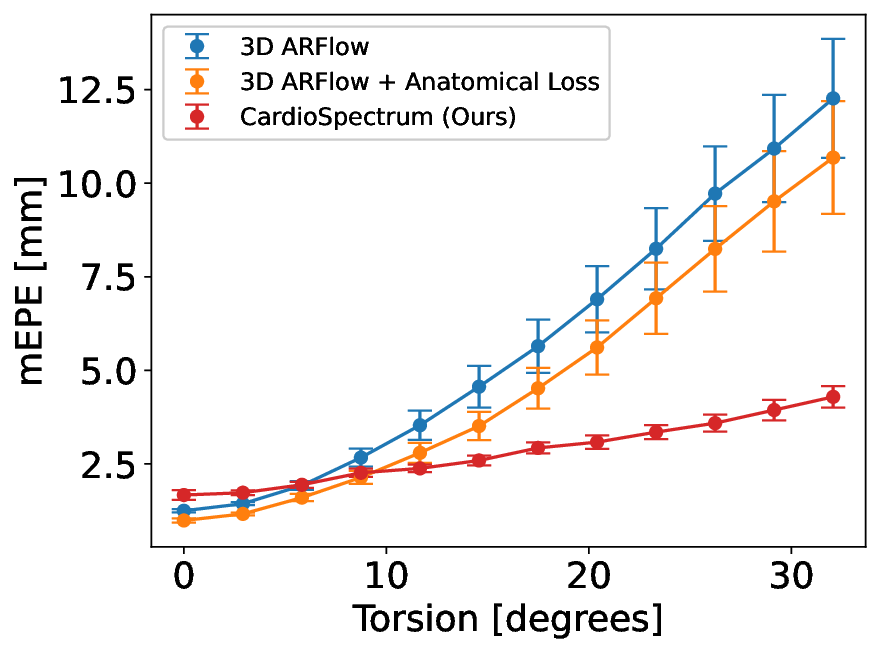}
        \caption{Overall\\mEPE}
        \label{fig:sub1}
    \end{subfigure}
    \begin{subfigure}{0.315\textwidth} 
        \includegraphics[width=\linewidth]{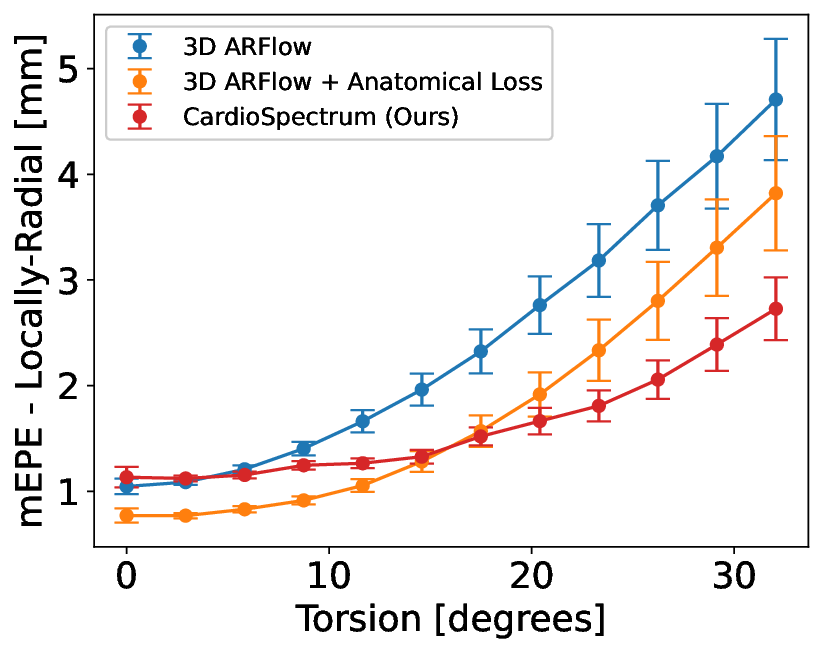}
        \caption{Locally-Radial\\mEPE}
        \label{fig:sub2}
    \end{subfigure}
    \begin{subfigure}{0.324\textwidth}
        \includegraphics[width=\linewidth]{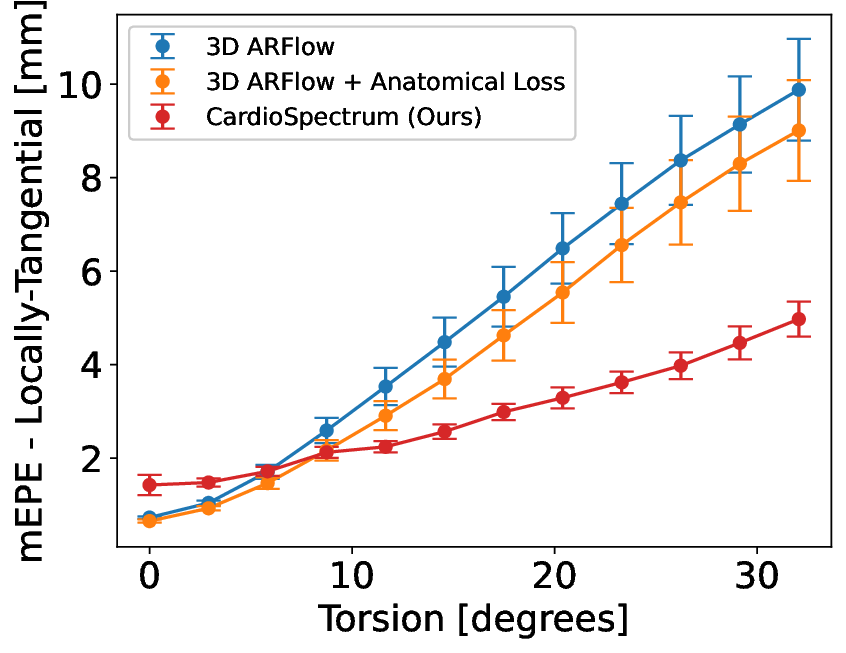}
        \caption{Locally-Tangential\\mEPE}
        \label{fig:sub3}
    \end{subfigure}    
    \caption{Comprehensive comparison of proposed model and two baselines in accurately capturing cardiac deformation across a range of torsion angles. 
    The metrics evaluated included the mEPE over the myocardium volume, the mEPE in locally-radial and locally-tangential directions within the segmentation hull. 
    As depicted, our model exhibited strengths that were particularly evident in \ref{fig:sub1} representing the overall mEPE, and pronounced advantages in locally-tangential components \ref{fig:sub3}, which aligns with our method’s goal of addressing the aperture problem, a challenge that was difficult to face by the two baseline models. Error bars represent the standard errors (SE). See Fig. 1 in supplementary material for the component-wise mEPE in radial, circumferential, and longitudinal directions and mean angular error.}
    \label{fig:main}
\end{figure}

Subfigure~\ref{fig:sub1} shows that as the torsion angle increased, the mEPE across the myocardium region in our method consistently exhibited a reduction in comparison to the baselines.
In terms of the locally-radial and locally-tangential flow components within the segmentation hull (Subfigures~\ref{fig:sub2},~\ref{fig:sub3}), our method demonstrated superior performance as the torsion angle increased, particularly in capturing the intricacies of locally-tangential flow patterns compared to the baseline models. Our model exhibited significant advantages in locally-tangential components, and its performance in locally-radial aspects still remained competitive, with some variations observed in lower torsion ranges. This emphasizes the ability of our approach to effectively capture locally-tangential flow patterns within the anatomical regions of interest, by addressing the challenges associated with the aperture problem.
Breaking down the mEPE into radial, circumferential, and longitudinal components provides additional insights. For detailed analyses on these components, as well as the mean angular error, see the supplementary material.

The overarching trend in our results thus indicates that our method excels with increasing torsion angles compared to the baseline models.

%% file: 06_discussion_and_conclusions.tex
Our comprehensive approach that integrates a 3D DL architecture with spectral methods makes a notable contribution to LV flow analysis. We conducted a wide range of experimental evaluations that show a reduction in mEPE across myocardial regions, particularly in scenarios with increased torsion complexity, thus highlighting the superiority of our approach over baseline methods.

One of the key strengths of our method is its ability to address locally-tangential flow components, thus effectively overcoming challenges associated with the aperture problem. The constraints loss, coupled with the integration of 2D constraints from FMs, contributes to greater accuracy in predicting optical flow and has strong implications for cardiac health and disease.

%% file: 09_supplemetary_material.tex
\begin{figure}
    \centering
    \begin{subfigure}{0.45\textwidth} 
        \includegraphics[width=\linewidth]{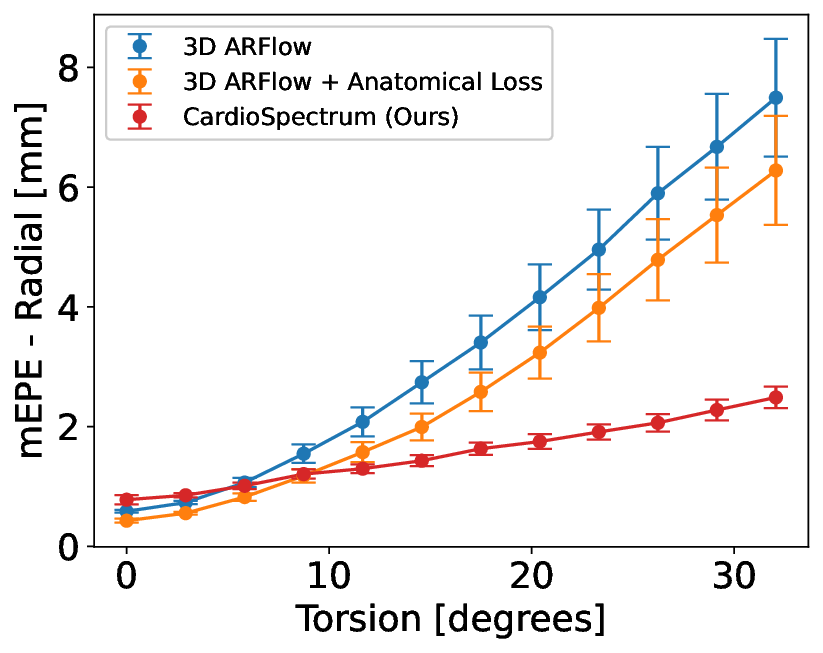}
        \caption{Radial mEPE}
        \label{fig:sub1}
    \end{subfigure}
    \begin{subfigure}{0.45\textwidth} 
        \includegraphics[width=\linewidth]{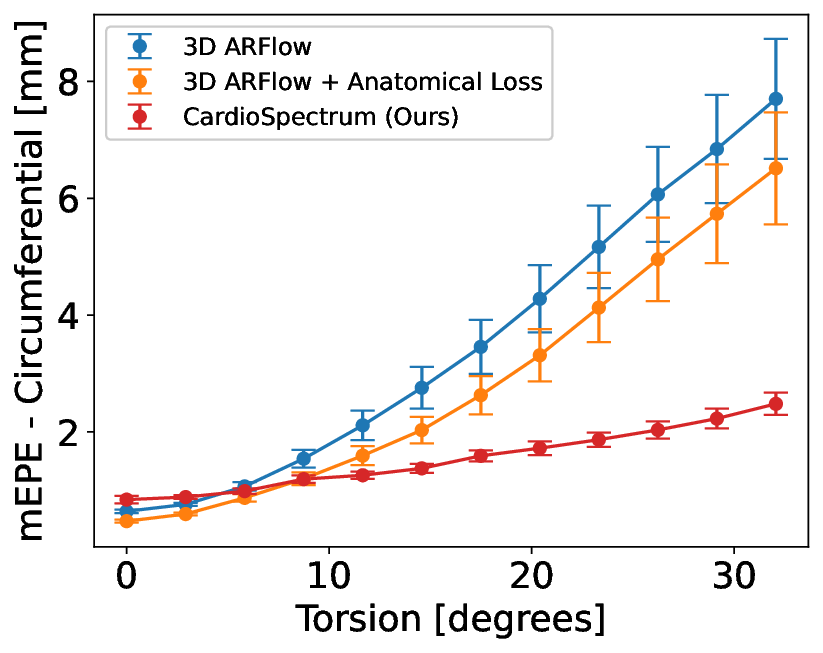}
        \caption{Circumferential mEPE}
        \label{fig:sub2}
    \end{subfigure}
    
    \begin{subfigure}{0.45\textwidth}
        \includegraphics[width=\linewidth]{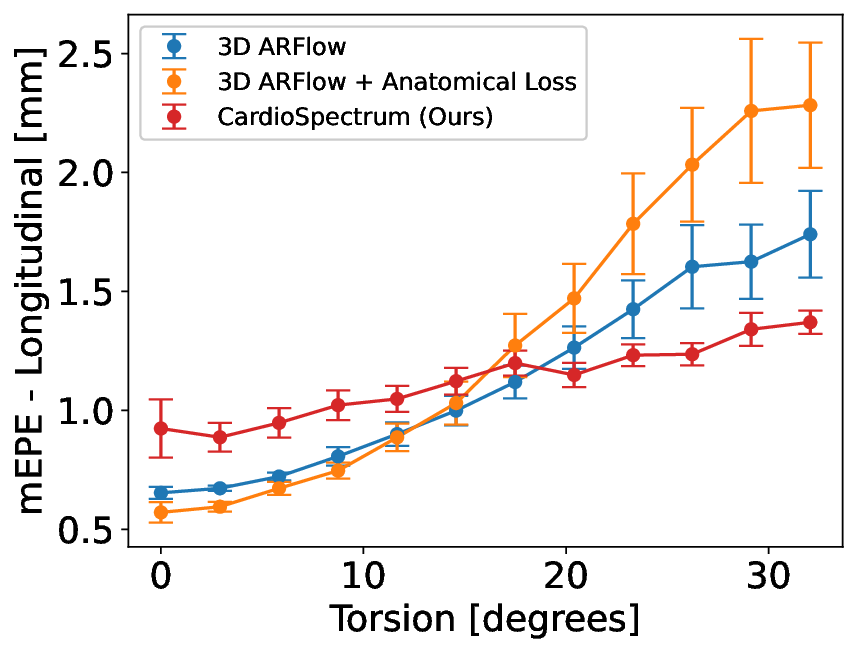}
        \caption{Longitudinal mEPE}
        \label{fig:sub3}
    \end{subfigure}
    \begin{subfigure}{0.45\textwidth} 
        \includegraphics[width=\linewidth]{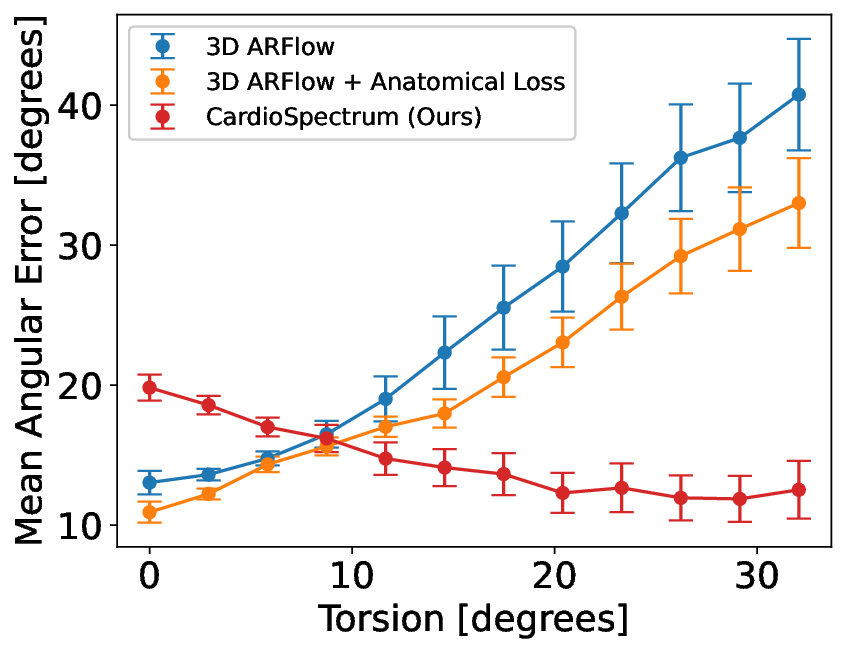}
        \caption{Angular Error}
        \label{fig:sub4}
    \end{subfigure}    
    \caption{Additional comparison of the proposed model and two baselines in capturing cardiac deformation across torsion angles, breaking down mEPE into radial (\ref{fig:sub1}), circumferential (\ref{fig:sub2}), and longitudinal (\ref{fig:sub3}) components. Radial and circumferential mEPE consistently showcase our model’s superior performance. Although performance relatively decreased at lower torsion angles in the longitudinal component, all models had lower errors than in other components. \\Subfigure~\ref{fig:sub4} shows dissimilarity between predicted and true flow directions based on angular error, with our method outperforming baselines particularly at higher torsion angles. Error bars represent standard errors (SE). These figures supplement Figure 3 and Section 5.1 of the main paper.
    }
    \label{fig:main}
\end{figure}

\begin{table}
\caption{Comparison of mEPE within the myocardium segmentation mask, along SE, at various torsion angles using CardioSpectrum and two baseline methods. Bold values denote significant improvement, and underlined values indicate comparable performance.}\label{tab1}
\begin{tabular}{p{2cm} p{2.5cm} p{2.5cm} p{3.5cm}}
\Xhline{1.5pt}  
\multirow{3}{*}{Torsion [deg]} & \multicolumn{3}{c}{mEPE $\pm$ SE [mm]} \\
\cmidrule(lr){2-4}
 & 3D ARFlow & \multirow{2}{*}{\shortstack[l]{3D ARFlow +\\Anatomical Loss}} & \multirow{2}{*}{\shortstack[l]{CardioSpectrum \\ (Ours)}} \\
 & & & \\
\midrule

0    & 1.25 $\pm$ 0.043 & \textbf{0.99 $\pm$ 0.06} & 1.67 $\pm$ 0.13 \\
2.9  & 1.44 $\pm$ 0.039 & \textbf{1.16 $\pm$ 0.04} & 1.73 $\pm$ 0.06 \\
5.8  & 1.92 $\pm$ 0.11 & \textbf{1.60 $\pm$ 0.10} & 1.94 $\pm$ 0.08 \\
8.7  & 2.67 $\pm$ 0.24 & \underline{2.14 $\pm$ 0.17} & \underline{2.26 $\pm$ 0.11} \\
11.7 & 3.53 $\pm$ 0.39 & 2.80 $\pm$ 0.27 & \textbf{2.38 $\pm$ 0.10} \\
14.6 & 4.56 $\pm$ 0.56 & 3.51 $\pm$ 0.38 & \textbf{2.59 $\pm$ 0.13} \\
17.5 & 5.65 $\pm$ 0.71 & 4.52 $\pm$ 0.55 & \textbf{2.93 $\pm$ 0.15} \\
20.4 & 6.90 $\pm$ 0.88 & 5.61 $\pm$ 0.72 & \textbf{3.08 $\pm$ 0.18} \\
23.3 & 8.25 $\pm$ 1.08 & 6.93 $\pm$ 0.96 & \textbf{3.35 $\pm$ 0.19} \\
26.2 & 9.72 $\pm$ 1.26 & 8.24 $\pm$ 1.14 & \textbf{3.59 $\pm$ 0.22} \\
29.1 & 10.93 $\pm$ 1.43 & 9.51 $\pm$ 1.34  & \textbf{3.94 $\pm$ 0.27} \\
32.1 & 12.26 $\pm$ 1.59 & 10.68 $\pm$ 1.50 & \textbf{4.29 $\pm$ 0.29} \\

\Xhline{1.5pt}  
\end{tabular}
\end{table}

\begin{figure}
    \centering
    \begin{subfigure}{0.32\textwidth} 
        \includegraphics[width=\linewidth]{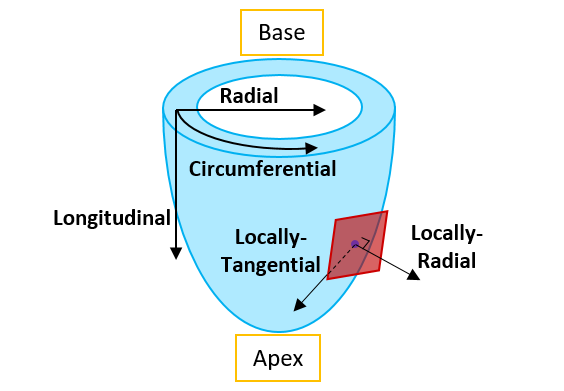}
        \caption{Coordinate System for Cardiac Movement}
        \label{fig:sub5}
    \end{subfigure}
    \begin{subfigure}{0.32\textwidth} 
        \includegraphics[width=\linewidth]{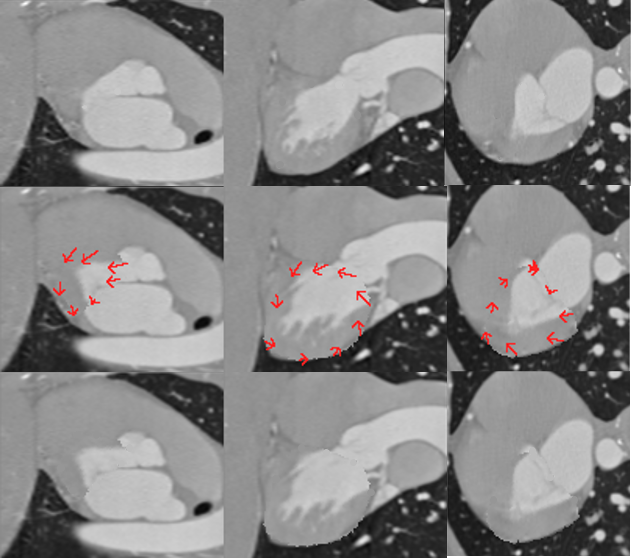}
        \caption{Visualization of Synthetic Deformation}
        \label{fig:sub6}
    \end{subfigure}
    \begin{subfigure}{0.34\textwidth} 
        \includegraphics[width=\linewidth]{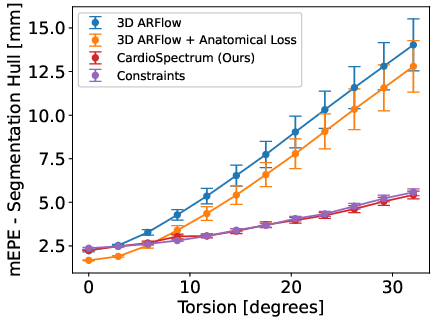}
        \caption{Segmentation Map Hull mEPE}
        \label{fig:sub7}
    \end{subfigure}
    \caption{
    (\ref{fig:sub5}) 
    The cardiac cycle involves \textit{radial} (LV cavity center to myocardial wall), \textit{circumferential} (tangential along the epicardial wall), and \textit{longitudinal} (along the LV’s long axis) LV movements. \textit{Locally-tangential} and \textit{locally-radial} movements refer to projections onto and perpendicular to the myocardial surface’s tangent plane.\\    
    (\ref{fig:sub6}) A sample before (top) and after (bottom) deformation. Arrows represent selected ground truth annotations. Sample from the 3D Slicer library (\url{http://www.slicer.org}).\\   
    (\ref{fig:sub7}) Comparing mEPE of CardioSpectrum, baselines, and ZoomOut-derived constraints across torsion angles, evaluated over the segmentation map hull. "Constraints" is composed of errors from ZoomOut and voxel-mesh conversions, impacting CardioSpectrum’s performance, especially at lower angles.}
    \label{fig:main2}    
\end{figure}